\newcommand{\dis}{\displaystyle} 
\documentstyle[12pt]{article}
\input epsf
\begin{document}
%%%%%%%%%%%%%%%%%%%%%%%
%%%% to get the nice 'draft' on each page
%%%\special{ !userdict begin /bop-hook{gsave 200 100 translate
%%%65 rotate /Times-Roman findfont 216 scalefont setfont
%%%0 0 moveto .98 setgray (DRAFT) show grestore}def end}
%%%%%%%%%%%%%%%%%%%%%%%
\parindent 1.3cm
\thispagestyle{empty}   % to suppress the page number on the first page
\vspace*{-3cm}
\noindent

\def\arccot{\mathop{\rm arccot}\nolimits}
\def\sd{\strut\displaystyle}

\begin{obeylines}
\begin{flushright}
UG-FT-57/96
hep-ph/9602401
February 1996
\end{flushright}
\end{obeylines}

\vspace{2cm}

\begin{center}
\begin{bf}
\noindent
CHIRAL PERTURBATION THEORY
\footnote{Talk presented at the International Conference on Particle Physics
and Astrophysics in The Standard Model and Beyond, Bystra (Poland), 
September 1995.}
\end{bf}
  \vspace{1.5cm}\\
FERNANDO CORNET
\vspace{0.1cm}\\
Departamento de F\'\i sica Te\'orica y del Cosmos,\\
Universidad de Granada, 18071 Granada, Spain\\
   \vspace{2.2cm}

{\bf ABSTRACT}
\end{center}

\parbox[t]{12cm}{\indent

An introduction to the methods and ideas of Chiral Perturbation
Theory is presented in this talk. The discussion is illustrated
with some phenomenological predictions that can be compared with
available experimental results.}

\newpage
\noindent
{\bf 1.-- Introduction}

Chiral Perturbation Theory (ChPT) is an effective theory that 
describes in a consistent and systematic way
the strong, electromagnetic and weak interactions involving
the lower mass pseudoscalar particles. It is based on a
non-proved theorem that states that apart from causality and
unitarity, the contents of a quantum field theory is dictated
by the symmetries it possesses \cite{WEINBERG}. The idea is, 
thus, to replace the quarks and gluons of QCD by the 
pseudoscalar mesons and write down the most general lagrangian
involving these particles that has the same symmetries as
the QCD lagrangian. The ChPT generating functional admits an
expansion in powers of external momenta and quark masses.
Although ChPT is not a renormalizable theory the results
can be rendered finite order by order in the expansion. 
The prize to pay is that new terms (with unknown constants)
have to be included in the lagrangian at each order in the expansion.

I cannot cover in a single talk all the exciting results related
with ChPT obtained during the last years. I should rather discuss here only 
a few of them and refer the interested reader to some of the more
recent, excellent reviews available in the literature
\cite{ECKER,PICH,BIJNENS,BEG,LEUTWYLER,BRAMON,MEISSNER}.

This talk is organized in the following way: In the next section I present
the lowest order lagrangian. In section 3, I present the way to calculate
the next order corrections, $O(p^4)$, and discuss some results at
$O(p^6)$. Section 4 deals with the Wess-Zumino term and its $O(p^6)$
corrections. Finally, the last section contains a brief summary of the
talk and a list (not an exhaustive one!) of the present active research
lines related with Chiral Perturbation Theory. 

\vglue 1.cm
\noindent
{\bf 2.-- Lowest Order Lagrangian.}
\vglue 0.5cm

The QCD lagrangian can be written in terms of $q= \hbox{column} (u \; d \; s)$
in the form:
\begin{equation}
\label{QCDL}
{\cal L}_{QCD} = - {1 \over 4} G_{\mu \nu} G^{\mu \nu}
                   + i \overline{q}_L \gamma^\mu D_\mu q_L
                   + i \overline{q}_R \gamma^\mu D_\mu q_R
                   + \overline{q}_L m_q q_R + \overline{q}_R m_q q_L 
 + {\cal L}_{QCD}^{HF},
\end{equation}
where the term ${\cal L}_{QCD}^{HF}$ includes the contribution from the
heavy quarks ($c$, $b$ and $t$) and we have explicitly separated the
contributions for the left-handed,$q_L$, and right-handed, $q_R$,
light quark fields. These components appear always separated except in
the two terms proportional to the quark mass matrix, $m_q = \hbox{diag}
(m_u \; m_d \; m_s)$.
It is, thus, clear that in the limit where $m_q = 0$
the lagrangian is invariant under independent transformations of the
left and right-handed quark fields,i.e. under 
the group $SU(3)_L \times SU(3)_R$:
\begin{equation}
\label{QTRANS}
q_L \to g_L q_L \qquad q_R \to g_R q_R \qquad \hbox{with} \qquad
g_L,g_R \in SU(3)_{L,R}.
\end{equation}
In view of this symmetry, one would expect all the hadrons to appear 
in multiplets of opposite parity, where all the particle should 
have approximately the same mass. However, there is no evidence for
a particle with the same quantum numbers as the proton, but opposite 
parity and similar mass (the lightest $I(J^P) = {1 \over 2}({1 \over 2}^-)$
state has a mass $m \sim 1535 \; MeV$). Similar comparisons can be made for
all the other hadronic states. We cannot blame the small
quark masses $m_u$ and $m_d$ for this big effect. Instead, before the
appearance of QCD it was already recognized that $SU(3)$ was a rather
good symmetry \cite{GN}. The chiral symmetry is, thus, spontaneously
broken:  $SU(3)_L \times SU(3)_R \to SU(3)_V$, i.e. the vacuum is symmetric
only under $SU(3)_V$ transformations. This breaking is produced through
the non-zero value of the quark condensate
\begin{equation}
\label{CONDENSATE}
<0|\overline{u}u|0> = <0|\overline{d}d|0> = <0|\overline{s}s|0>
\sim -(250 \; MeV)^{3/2},
\end{equation}
which becomes the order parameter of the spontaneous chiral symmetry breaking.
The Goldstone theorem assures that in this process eight goldstone bosons
appear (one for each broken generator) \cite{GOLDSTONE}. These bosons are
massless in the limit of massless quarks, but the small explicit chiral
symmetry breaking through the quark masses gives a small mass
to the goldstone bosons. 

The idea of ChPT is to write down an effective lagrangian where the quarks
and gluons have been replaced by the goldstone bosons appearing in the
spontaneous chiral symmetry breaking. A convenient parameterization is in
terms of a $3 \times 3$ unitary matrix:
\begin{equation}
\label{SIGMA}
\Sigma = e^{2iM/f}   \qquad \hbox{with} \qquad
M = \left( \begin{array}{ccc}
   \dis{\pi^0 \over \sqrt{2}} + \dis{\eta \over \sqrt{6}} & \pi^+ & K^+ \\
    \pi^- & -\dis{\pi^0 \over \sqrt{2}} + \dis{\eta \over \sqrt{6}} & K^0 \\
    K^- & \overline{K}^0 & - 2 \dis{\eta \over \sqrt{6}}
    \end{array} \right) ,
\end{equation}
and $f$ is a free constant. This matrix transforms under 
$SU(3)_L \times SU(3)_R $ as:
\begin{equation}
\label{TRANSFORMATION}
 \Sigma \to g_L \Sigma g_R^\dagger .
\end{equation} 
The effective lagrangian contains an infinite number of terms, but it can be
expanded according to the number of derivatives. This is something more than 
a convenient classification. Physically, it means an expansion in terms of
powers of momenta that have to be small compared with the chiral symmetry 
breaking scale, which is $\sim 1 \; GeV$. Lorentz invariance requires 
the number of derivatives to be even. Thus, the first term is:
\begin{equation}
\label{LO1}
{\cal L}_2 = {f^2 \over 8} tr \partial_\mu \Sigma \partial^\mu \Sigma^\dagger.
\end{equation}
This is the only relevant term with two derivatives, because other 
possible terms
one can think off, such as $\Sigma \partial_\mu \partial^\mu \Sigma^\dagger$,
differ from (\ref{LO1}) only in a total derivative. Expanding $\Sigma$ it is
obvious that the lagrangian in
Eq. (\ref{LO1}) contains the kinetic terms for all the pseudoscalar
mesons and interaction terms involving $4$, $6$ and a larger number of
pseudoscalars. Moreover, taking the axial current one has:
\begin{equation}
\label{AXIAL}
<0|J_\mu^{L_{1+i2}}|\pi^+> = -\dis{i \over \sqrt{2}}f_\pi P_{\pi_\mu}
\quad \hbox{with} \quad J_\mu^{L_a} = - \dis{if^2 \over 4} 
tr(T^a \partial_\mu \Sigma \Sigma^\dagger),
\end{equation}
leading to the identification at this order of the free constant $f$ with the
well-known pion decay constant $f_\pi = f =132 \; MeV$. Note that at this
point there is a complete $SU(3)$ symmetry among the three decay constants:
$f_\pi = f_\eta = f_K$.

The effects of the explicit chiral symmetry breaking through the 
non-vanishing values of the quark masses can be included in the lagrangian
(\ref{LO1}) adding some new terms:
\begin{equation}
\label{LO2}
{\cal L}_2 = {f^2 \over 8} tr \left(
               \partial_\mu \Sigma \partial^\mu \Sigma^\dagger
           + ( \Sigma \chi^\dagger + \chi \Sigma^\dagger) \right),
\end{equation}
where $\chi$ contains the external scalar and pseudoscalar fields in the 
following way:
\begin{equation}
   \label{chi}
 \chi = B (s - ip) , \qquad \hbox{where} \qquad s = m_q + \cdots \quad .
\end{equation} 
$B$ is again a free constant that can be calculated in terms of the
pseudoscalar and quark masses:
\begin{equation}
\label{B}
B = \dis{2 m_\pi^2 \over m_u+m_d} 
  = \dis{2 m_K^2 \over m_u+m_s} 
  = \dis{6 m_\eta^2 \over m_u+m_d+m_s}. 
\end{equation}
From this relations, eliminating the quark masses, one can obtain the 
Gell-Mann-Okubo mass relation \cite{GM}
\begin{equation}
\label{GELL-MANN_OKUBO}
4 m_K^2 - m_\pi^2 = 3 m_\eta^2.
\end{equation}
The new term in the lagrangian also contains more interaction terms, which
are proportional to the pseudoscalar masses. The expansion, thus, is not
only in powers of the momenta, but also in powers of the pseudoscalar masses.

External vector fields can be introduced in the theory converting the
derivatives appearing in the lagrangian in covariant derivatives:
\begin{equation}
\label{LO}
\begin{array}{c}
{\cal L}_2 = \dis{f^2 \over 8} tr (D_\mu \Sigma D^\mu \Sigma^\dagger
           + ( \Sigma \chi^\dagger + \chi \Sigma^\dagger)); \\ \quad \\
D_\mu \Sigma = \partial_\mu \Sigma + i L_\mu \Sigma - i \Sigma R_\mu
\end{array}
\end{equation}
and adding the appropriate kinetic terms for the vector fields $L_\mu$ and
$R_\mu$. These fields transform under $SU(3)_L \times SU(3)_R $ as:
\begin{equation}
\label{VTRANSFORMATIONS}
\begin{array}{c}
L_\mu \to g_L L_\mu g_L^\dagger -i g_L \partial_\mu g_L^\dagger , \\ \quad \\
R_\mu \to g_R R_\mu g_R^\dagger -i g_R \partial_\mu g_R^\dagger .
\end{array}
\end{equation}
In particular, we can introduce electromagnetic interactions involving 
photons and pseudoscalars with the identification
$L_\mu = R_\mu = e A_\mu Q$, where $Q$ is the quark charge matrix:
\begin{equation}
\label{Q}
Q = 
\left( \begin{array}{ccc}
       \dis{2 \over 3}  &          0         &         0         \\
             0          &  -\dis{1 \over 3}  &         0         \\
             0          &          0         &  -\dis{1 \over 3}  
       \end{array}  \right).
\end{equation}  

In this way we complete the description of the lowest order ChPT lagrangian.
With this lagrangian we can reproduce all the Current Algebra results
obtained in the 60's. For instance, it is a very simple exercise to 
obtain from Eq. (\ref{LO}) the Weinberg amplitude \cite{WEINBERG2}:
\begin{equation}
  \label{PIPILO}
A(s,t,u) = {s - m_\pi^2 \over f_\pi^2},
\end{equation}
that fixes the scattering amplitude for the process
$\pi^a(p_a) \pi^b(p_b) \to \pi^c(p_c) \pi^d(p_d)$ through the isospin
decomposition
\begin{equation}
T_{ab,cd} = \delta_{ab} \delta_{cd} A(s,t,u)
           +\delta_{ac} \delta_{bd} A(t,s,u)
           +\delta_{ad} \delta_{bc} A(u,t,s),
\end{equation}
with $s=(p_a+p_b)^2$, $t=(p_a-p_c)^2$ and $u=(p_a-p_d)^2$. The amplitudes
of definite isospin can be expanded in partial wave amplitudes according
to:
\begin{equation}
A^I(s, \cos \theta) = i \dis{32 \pi \sqrt{s} \over \sqrt{s-4m_\pi^2}}
      \sum_{l=0}^{\infty} (2l+1) P_l(\cos \theta)
       (1 - e^{2i\delta_l^i(s)}),
\end{equation}
where $\delta_l^i$ are the phase shifts. The corresponding scattering lengths,
$a_l^i$, 
are defined as the slope of the phase shifts at threshold. The lowest order
predictions from Eq. (\ref{PIPILO}) are:
\begin{equation}
a_0^0 = 0.156 \qquad \qquad \qquad a_0^0 - a_2^0 = 0.201
\end{equation}
to be compared with the experimental results:
\begin{equation}
a_0^0 = 0.26 \pm 0.05 \qquad \qquad \qquad  a_0^0 - a_2^0 = 0.29 \pm 0.04.
\end{equation}
It is clearly important to evaluate what are the corrections to these lowest
order results.

\vglue 1.cm
\noindent
{\bf 3.-- Higher Order Corrections.}
\vglue 0.5cm

The advantage of ChPT is that it provides a consistent way to calculate
the quantum corrections to the tree level results of Current Algebra.
The key point is that loop diagrams always contribute to a higher
order in the momentum expansion. For instance, one loop diagrams with
vertices derived from the lagrangian ${\cal L}_2$ are $O(p^4)$. In any 
one-loop diagram the number of vertices is the same as the number of
internal lines. Since each internal line contributes at $O(p^{-2})$,
the total dimension of the diagram is given by the momentum integral,
i.e. $O(p^4)$. This result can be easily generalized to any $L$-loop
diagram containing $N_d$ vertices of dimension $d$
\begin{equation}
  \label{COUNTING}
D=2L +2 + \sum_{d} (d-2)N_d.
\end{equation}

The result of a loop calculation is, in general, divergent as one expects 
from dimensional counting. However, consistency in the momentum expansion
requires the introduction of the terms of the effective lagrangian that
are of the same order as the loop result. These terms are multiplied by
free constants. So, we can use these constants to remove all the
divergences, just splitting all of them into a finite, renormalized piece
and an infinite one that is tuned to absorb all the divergences appearing
at a given order in the expansion. Since we have built the effective 
lagrangian in such a way that contains {\it all} the possible terms at each
order, we are assured that we will be able to
remove all the divergences. The theory, however, is non-renormalizable
because we are forced to introduce new counterterms at each
order in the chiral expansion, in contrast to a renormalizable theory
where only a finite number of counterterms are needed. We should remark here
that both, the finite part of the constants and the loop contributions depend 
on the renormalization scale $\mu$. The physical amplitudes, however,
are independent of this scale.

The lagrangian at $O(p^4)$ contains $10$ terms contributing to the
same processes as the lagrangian ${\cal L}_2$ \cite{GL}. In addition 
there are two more terms that do not contain any pseudoscalar field and,
thus, they cannot be measured. The values of the $10$ free constants can be 
determined from experimental data \cite{GL,BIJNENS}. Their values at
the scale $\mu = m_{\rho}$, together with an indication of
the process where they have been determined is shown in Table 1. Actually, 
they turn out to be of the expected order of magnitude. Indeed, assuming
that the chiral symmetry breaking scale, $\Lambda_\chi$, is $O(1 \; GeV)$
and taking into account that the constant in the lagrangian ${\cal L}_2$
is $f_\pi/8$, we would expect
\begin{equation}
L_i \sim {f_\pi \over 8 \Lambda_\chi} = 2 \times 10^{-3}.
\end{equation}

\begin{table}
\begin{center}
\begin{tabular}{|c|c|c|}
\hline
$L_i$ & Value $\cdot 10^3$ & Input\\
\hline
1 & $0.4\pm0.3 $& $K_{e4}$ and $\pi\pi\to\pi\pi$\\
2 & $1.35\pm0.3 $& $K_{e4}$ and $\pi\pi\to\pi\pi$\\
3 & $-3.5\pm1.1$ & $K_{e4}$ and $\pi\pi\to\pi\pi$\\
4&$-0.3\pm0.5$&$1/N_c$ arguments\\
5&$1.4\pm0.5$&$F_K/F_\pi$\\
6&$-0.2\pm0.3$&$1/N_c$ arguments\\
7&$-0.4\pm0.2$&Gell-Mann-Okubo, $L_5$, $L_8$\\
8&$0.9\pm0.3$&$m_{K^0}-m_{K^+}$, $L_5$, baryon mass ratios\\
9&$6.9\pm0.7$&pion electromagnetic charge radius\\
10&$-5.5\pm0.7$&$\pi\to e\nu\gamma$\\
\hline
\end{tabular}
\end{center}
\caption{The values of the $L_i$ coefficients and the input used
to determine them, they are quoted at a scale $\mu=m_\rho$.}
%\rlabel{table1}
\end{table}

A first result at $O(p^4)$ is the $SU(3)$ breaking in the decay constants:
\begin{equation}
  \label{F}
\begin{array}{l}
f_\pi = f \left[ 1 - 2\mu_\pi - \mu_K + \dis{4 m_\pi^2 \over f^2} L_5(\mu)
               + \dis{8 m_K^2 + 4 m_\pi^2 \over f^2} L_4(\mu) \right] \\
\quad \\
f_K = f \left[ 1 - \dis{3 \over 4} \mu_\pi - \dis{3 \over 2} \mu_K 
             - \dis{3 \over 4} \mu_{\eta_8} 
             +  \dis{4 m_K^2 \over f^2} L_5(\mu)
             + \dis{8 m_K^2 + 4 m_\pi^2 \over f^2} L_4(\mu) \right] \\
\quad \\
f_{\eta_8} = f \left[ 1 - 3 \mu_K 
             +  \dis{4 m_{\eta_8}^2 \over f^2} L_5(\mu)
             + \dis{8 m_K^2 + 4 m_\pi^2 \over f^2} L_4(\mu) \right] ,
\end{array}
\end{equation}
where $\mu_P$ arises from the loop contributions and is given by:
\begin{equation}
\mu_P = \dis{m_P^2 \over 16 \pi^2 f^2} 
                           \log \left(\dis{m_P^2 \over \mu^2 }\right).
\end{equation} 
The ratios among the decay constants are almost independent of the
value of $L_4$, which is expected to be zero because it is of
higher order in the $1/N_c$ expansion. Thus, from the experimental value
\begin{equation}
\dis{f_K \over f_\pi} = 1.22 \pm 0.01
\end{equation}
we can fix the constant $L_5$ to the value quoted in Table 1 and predict
the ratio
\begin{equation}
\dis{f_{\eta_8} \over f_\pi} = 1.30 \pm 0.05.
\end{equation}

The constants in the lagrangian can, a priori, be calculated from QCD but
certainly a non-perturbative method is required. There are some 
interesting attempts to investigate the chiral lagrangian with lattice
QCD but, up to now the quenched approximation has always been used and,
thus, a direct comparison with the numbers in Table 1 is meaningless and
a reformulation of the chiral lagrangian in the quenched approximation
is needed \cite{GOLTERMAN}. A different approach to evaluate the $L_i$
was adopted in \cite{ECKER2,DONOGHUE}. The assumption was that the constants
$L_i$ are saturated by the contribution of the lowest mass resonances
after they have been integrated out. The predicted values of the 
constants under this assumption is shown in Table 2. Since the couplings
of the resonances to the pseudoscalar mesons are also unknown constants
we have to use three of the $L_i$ constants to fix these couplings. In
any case, we can see from the table that the agreement with the experimental
values is excellent. Although it is not shown in the table it can be seen
that, as one would expect, the dominant contribution arises from the
vector meson nonet. This assumption is particularly useful when calculating
$O(p^6)$ corrections, where the number of free constants is very large to be
fixed by experimental data and strict (not implemented with this assumption)
Chiral Perturbation Theory looses predictive power.

\begin{table}
\begin{center}
\begin{tabular}{|c|c|c|}
\hline
$L_i$ & Value $\cdot 10^3$ & Resonance Saturation Prediction\\
\hline
1 & $0.4\pm0.3 $& $\phantom{-}0.6\phantom{^*}$\\
2 & $1.35\pm0.3 $& $\phantom{-}1.2\phantom{^*}$\\
3 & $-3.5\pm1.1$ & $-3.0\phantom{^*}$\\
4&$-0.3\pm0.5$&$\phantom{-}0.0\phantom{^*}$\\
5&$1.4\pm0.5$&$\phantom{-}1.4^*$\\
6&$-0.2\pm0.3$&$\phantom{-}0.0\phantom{^*}$\\
7&$-0.4\pm0.2$&$-0.3\phantom{^*}$\\
8&$0.9\pm0.3$&$\phantom{-}0.9^*$\\
9&$6.9\pm0.7$&$\phantom{-}6.9^*$\\
10&$-5.5\pm0.7$&$-6.0\phantom{^*}$\\
\hline
\end{tabular}
\end{center}
\caption{The values of the $L_i$ coefficients compared with the predictions
obtained assuming their saturation by the resonances. The asterisks mark the
input parameters.}
%\rlabel{table1}
\end{table}

Before closing this section let me comment on two recent two loop
calculations. We showed in the previous section that the lowest order
prediction for the scattering length $a_0^0$ was slightly out of the
experimental value. The one-loop, $O(p^4)$ correction was calculated
long ago \cite{GL2,DONOGHUE2,GASSER} and recently the two-loop calculation
has been performed \cite{BIJNENS3}. We show in Fig. 1 the phase shift
difference $\delta_0^0 - \delta_1^1$ as a function of the center of
mass energy of the two incoming pions. The contribution from the
constants in ${\cal L}_6$ has been estimated to be negligible and, thus,
the figure has been drawn assuming that they cancel. The scattering lengths
also show the same improvement:
\begin{equation}
  \begin{array}{l}
a_0^0 = 0.156 + 0.044 + 0.017 = 0.217 \\
\quad \\
a_0^0 - a_0^2 = 0.201 + 0.042 + 0.016 = 0.258,
\end{array}
\end{equation}
where the first term in the addition correspond to the lowest order result, 
the second to the one-loop correction and the third to the 
two-loop correction.

The process $\gamma \gamma \to \pi^0 \pi^0$  presents a very interesting
situation. Inspecting the lagrangians ${\cal L}_2$ and ${\cal L}_4$ one
can see that there are no $\gamma \pi^0 \pi^0$ nor 
$\gamma \gamma \pi^0 \pi^0$ interaction terms. Thus, the lowest order 
contribution to $\sigma(\gamma \gamma \to \pi^0 \pi^0)$ is given by
the loop diagrams shown in Fig. 2, where the particles circulating in the 
loops are charged pions and kaons. This is an $O(p^4)$ contribution.
Thus, the result from the loop calculation must be finite (because there
are no terms in ${\cal L}_4$ contributing to this process that can be
used to remove the divergences). Indeed, although each one of the
diagrams shown in Fig. 2 is divergent, when adding all of them the 
divergences cancel and we obtain a parameter free prediction for the
cross-section. The result is shown in Fig. 3 (dashed line) \cite{BC}
compared with the experimental data from the Crystall Ball Collaboration
\cite{CBC}. Although the order of magnitude of the theoretical
prediction is correct, it is a factor $\sim 2$ too small compared with
the experimental data near threshold. The inclusion of the $O(p^6)$ terms
improves the agreement between the theoretical prediction and the
experimental data \cite{BGS} (solid line). These corrections receive 
contributions from two-loop diagrams constructed with vertices from 
${\cal L}_2$ , one loop diagrams with one vertex from ${\cal L}_4$
and tree level contributions from ${\cal L}_6$. Again, these last 
contributions contain free parameters that have fixed 
assuming their saturation by resonances.

\begin{figure}
\setlength{\unitlength}{1cm}
\begin{picture}(19.,6.)
\epsfxsize=18cm
\put(-1.5,-10.){\epsfbox{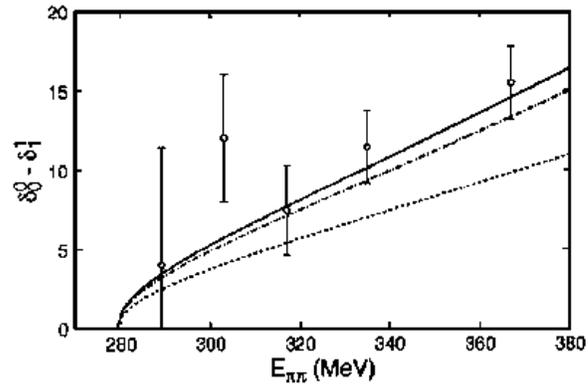}}
\end{picture}
\caption{ Phase shift difference as a function of the center of mass energy.
The dashed line stands for the lowest order result, the dash-dotted line
for the one-loop result and the full line for the two-loops result, assuming
that the constants of the $O(p^6)$ lagrangian vanish.} 
\end{figure}

\begin{figure}
\setlength{\unitlength}{1cm}
\begin{picture}(13.,6.)
\epsfxsize=12cm
\put(2.,0.){\epsfbox{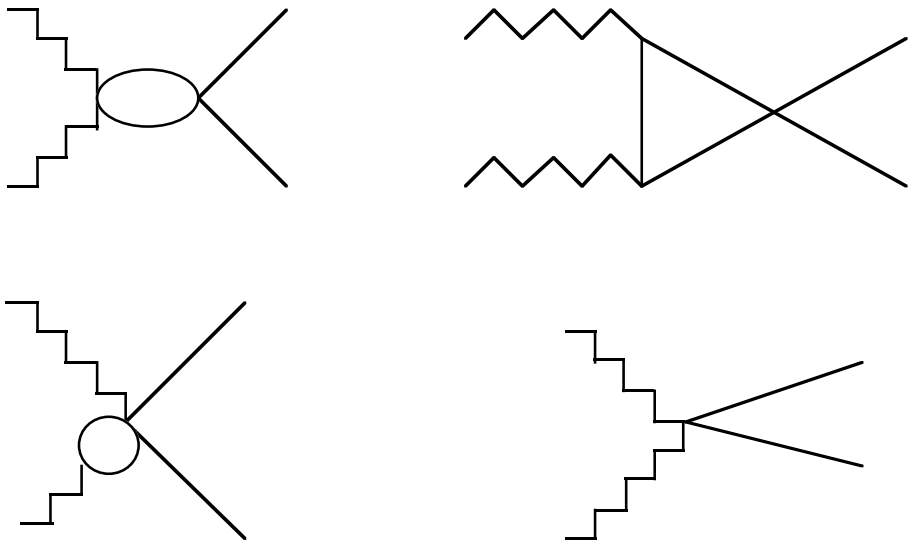}}
\end{picture}
\caption{Feynman diagrams contributing at lowest order to 
the process $\gamma \gamma \to \pi^0 \pi^0$}
\end{figure}

\begin{figure}
\setlength{\unitlength}{1cm}
\begin{picture}(17.,6.)
\epsfxsize=16cm
\put(0.5,-8.){\epsfbox{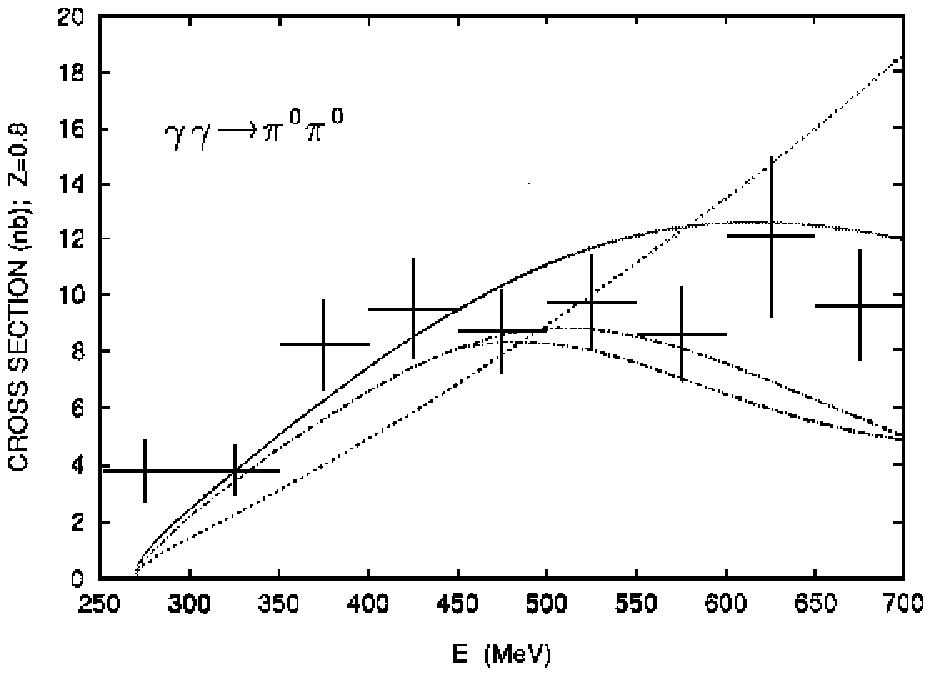}}
\end{picture}
\caption{Total cross section for the lowest order (dashed line) and $O(p^6)$
(solid line). The dash-dot band is a dispersive calculation from Ref. [25]}
\end{figure}

\vglue 1.cm
\noindent
{\bf 4.-- The Wess-Zumino Term and Higher Order Corrections.}
\vglue 0.5cm

The lagrangian at $O(p^4)$ contains an additional term originated by the
chiral anomaly: the Wess-Zumino term \cite{WZ}. This term contributes
to processes with an odd number of pseudoscalar fields, in contrast with
the lagrangians we dealt with in the previous sections that contribute
to processes with an even number of pseudoscalars. The most characteristic
process is the decay $\pi^0 \to \gamma \gamma$. Indeed, it is this process
the one that is used to fix the constant in the Wess-Zumino term to be
the number of colors \footnote{Due to the origin of the Wess-Zumino term,
the free constant it contains must be an integer \cite{WITTEN}}.

The Wess-Zumino term provides a good description of the decay width
$\Gamma ( \pi^0 \to \gamma \gamma)$ assuming
$f = f_\pi$. The situation, however, is rather different for the decay
$\eta \to \gamma \gamma$. The physical $\eta$ is a mixture of the octet and
singlet pieces:
\begin{equation}
  \label{ETMI}
    \begin{array}{l}
\eta = \cos \theta \eta_8 + \sin \theta \eta_1 \\
\eta^\prime = -\sin \theta \eta_8 + \cos \theta \eta_1
     \end{array}
\end{equation}
with $\theta = -19.5^o$ \cite{GILMAN}. Fixing $f_1 = 1.1 f_\pi$ from
the experimental value $\Gamma (\eta^\prime \to \gamma \gamma) =
(4.47 \pm 0.39) keV$ and using $f_8 = 1.3 f_\pi$, as predicted by
ChPT at $O(p^4)$, we obtain $\Gamma (\eta \to \gamma \gamma) = 0.44 keV$.
This result is in very good agreement with the experimental value
$\Gamma (\eta \to \gamma \gamma)_{exp} =(0.41 \pm 0.07) keV$. However,
the use of the next to leading order prediction for $f_8$ is inconsistent
with a lowest order prediction!. In order to take $f_8 \neq /f_\pi$
in a consistent way one has to include the whole, next order, $O(p^6)$
corrections. This was done in Refs. \cite{BBC1} and \cite{DONOGHUE3}, where
the explicit cancellation of all the divergences anid the absence of
contributions from the $O(p^6)$ lagrangian was shown. The only effect of
the next order corrections is the modification of the value of $f_8$, thus
justifying the procedure followed to obtain the ChPT prediction for the
$\eta \to \gamma \gamma$ decay width.

The cancellation of the corrections to the Wess-Zumino term is not a general
feature. In \cite{BBC1} it was explicitly shown that the cancellation
of the divergences appearing in one-loop diagrams contributing to the
process $P \to \gamma \gamma^*$ (where $P$ stands for a neutral pseudoscalar
meson and $\gamma^*$ is an off mass shell photon) requires the introduction
of counterterms. The $O(p^6)$ lagrangian contributing to anomalous
processes and the coefficients needed to cancel all the divergences are known
\cite{BBC2,AKHOURY,FEARING}. The number of terms in the lagrangian is 
again very large to determine them experimentally. However, with the
assumption of their saturation by the contribution of the lowest-mass
resonances a very good description of the $q^2$-dependence in the decay
$P \to \gamma \gamma^*$ has been achieved \cite{ABBC1}.

The $O(p^6)$ corrections clearly improve the situation for the decay
$\eta \to \pi^+ \pi^- \gamma$. The lowest order prediction
for the decay width turns out smaller than the experimental value:
\begin{equation}
 \label{ETAEX}
\begin{array}{lll}
\Gamma (\eta \to \pi^+ \pi^- \gamma)_{LO} = 35 \; eV & \qquad \qquad &
\Gamma (\eta \to \pi^+ \pi^- \gamma)_{EXP} = (53 \pm 10) \; eV.
\end{array}
\end{equation}
Moreover, the predicted photon energy spectrum does not fit the experimental
one \cite{LAYTER}. The $O(p^6)$ corrections have two effects. First of all,
they increase the value of the decay width to 
\begin{equation}
\Gamma (\eta \to \pi^+ \pi^- \gamma)_{O(P^6)} = 47 eV
\end{equation}
in such a way that it is now compatible with the experimental value
(\ref{ETAEX}). Second, they soften the photon spectrum as it is required 
by the experimental data.

Finally, a short comment on a case where the effects of the corrections to
the Wess-Zumino term are extremely important: the cross section for 
$\gamma \gamma \to \pi^0 \pi^0 \pi^0$. The lowest order amplitude turns
out to be proportional to $m_\pi^2$, thus predicting a very small 
cross section. At $O(p^6)$ terms proportional to the center of mass
energy squared appear giving rise to huge corrections that increase the
lowest order predictions in two orders of magnitude for a center of mass
energy around $600 \; GeV$ as can be seen in Fig. 4 \cite{TABBC}. 
Certainly, questions about the convergence of the expansion arise at
this point. However, there should be no problem because the huge corrections
are due to the smallness of the lowest order prediction. Indeed, in the
chiral limit, the lowest order amplitude vanishes, while the $O(p^6)$ 
contribution is different from $0$.

\begin{figure}
\setlength{\unitlength}{1cm}
\begin{picture}(13.,11.)
\epsfxsize=12cm
\put(2.,-4.){\epsfbox{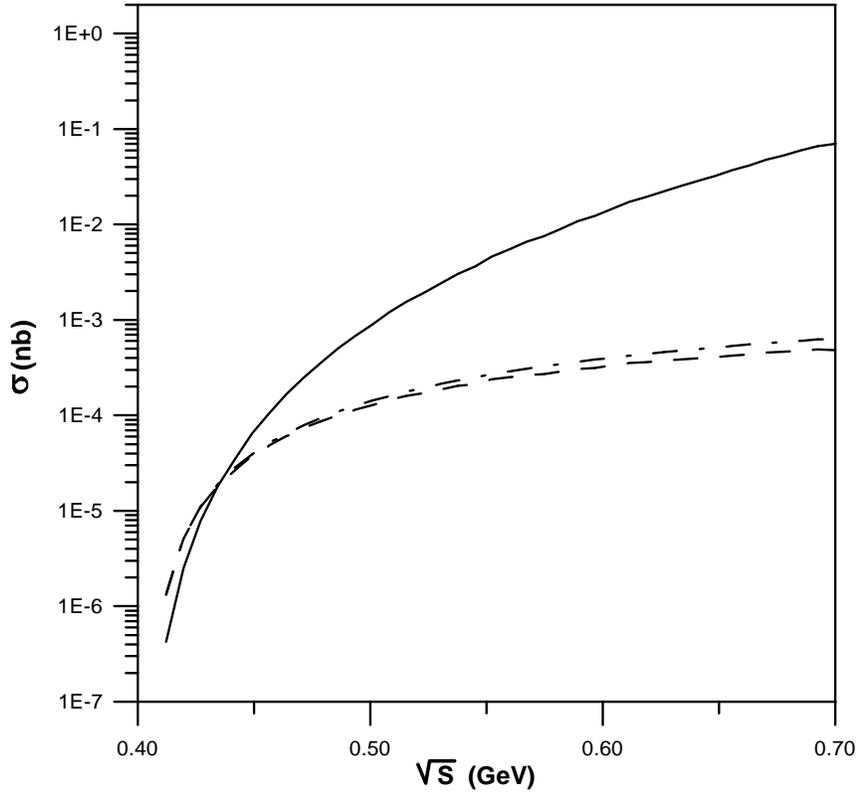}}
\end{picture}
\caption{$\gamma \gamma \to \pi^0 \pi^0 \pi^0$ cross section at lowest order
(dashed line) and $O(p^6)$ (full line) as a function of the center of mass
energy. The dot-dash line corresponds to the non-relativistic tree level
approximation.}
\end{figure}

\vglue 1.cm
\noindent
{\bf 5.-- Conclusions.}
\vglue 0.5cm

Chiral Perturbation Theory is an effective, low energy theory of QCD.
It allows to calculate cross-sections and decay widths for processes
involving pseudoscalar mesons. The expansion parameter is the momenta
and masses involved in the process compared to the chiral symmetry
breaking scale (around $1 \; GeV$). It is a non-renormalizable theory,
but results can be rendered finite order by order in the perturbative
expansion. The price to pay is the introduction of new free constants
in each order. This fact certainly  limits the predictivity
of the theory.
However, this is not important until a high order is reached, $O(p^6)$.
But even in this case, interesting phenomenological results
can be obtained
assuming the saturation of the free constants by the low mass
resonances.
The validity of this assumption has been verified for those constants
in the Chiral Lagrangian that can be fixed by experimental data.

Let me finish with a small list of research subjects in Chiral Perturbation
Theory that are being followed nowadays:
\begin{enumerate}
\item Two-loop calculations in the meson sector.
\item Introduction of Vector Mesons and other resonances in the Chiral 
lagrangian.
\item Kaon leptonic and non-leptonic decays.
\item Pion-nucleon interactions.
\item Chiral symmetry and underlying quark models.
\item Heavy Quark applications of chiral symmetry.
\end{enumerate}

\rightskip=0pc
\leftskip=0pc
\vglue 0.8cm
{\bf\noindent Acknowledgements \hfil}
\vglue 0.3cm

I thank A. Grau for her careful reading of the manuscript and M. Zralek
and J. Sladkowski for their kind hospitality during this workshop. This
work eas partially supported by CICYT under contract AEN 94-0936 and the
European Commision under contract CHRX-CT-92-0004.

{\bf\noindent References \hfil}
\vglue 0.3cm

\end{document}